\documentstyle[12pt]{article}

\renewcommand{\thefootnote}{\fnsymbol{footnote}}

\begin{document}

\begin{center}
{\Large $\eta-^4$He Bound States in the Skyrme Model}

\vspace{1cm}

{\large N. N. Scoccola$^{1,}\!\!$
\footnote[2]{Fellow of the CONICET, Argentina.} 
and D.O. Riska$^2$}

\end{center}

{\it 
$^1$Physics Department,  CNEA, 
Av. Libertador 8250, (1429) Buenos Aires, Argentina.}
\centerline{\it $^2$Department of Physics, University of Helsinki,
00014 Finland.}

\vspace{.7cm}
\centerline{July 1998}
\vspace{.7cm}

\begin{center}
{\large Abstract}
\end{center}

\begin{quotation}
The rational map ansatz for light nuclei in the Skyrme model is shown 
to imply the existence of an $\eta-^4$He bound state, with a binding 
energy of $\sim$~30~KeV.
\end{quotation}

\vskip 1.cm


\renewcommand{\thefootnote}{\arabic{footnote}}
\setcounter{footnote}{0}

\vspace{1cm}

Considerable progress has been made recently in the development 
of calculational methods for realistic applications of Skyrme's
topological soliton model for baryons \cite{Skyrme} to the 
structure of light nuclei \cite{Braaten,Sutcliffe,Houghton,Irw98}. 
This significantly extends the applicability of the Skyrme model,
which hitherto 
had largely been restricted to the structure of
baryons \cite{Zahed,Nyman}. The ``rational map'' method \cite{Houghton} 
in particular facilitates the development of simple yet realistic 
approximations to the full Skyrme model solution for spatially extended 
baryonic systems with A$>$1. A demonstration of the
utility of this method has recently been given in
the study of 
strange multiskyrmions in ref.\cite{SS98}.

We here apply the rational map ansatz to show that
$\eta$ mesons are very likely to form bound states with
$\alpha-$particles. This gives further support to the
long standing expectation that such should be formed
\cite{Bhal,Haider}, and which was reinforced by the
recent observation that the $\eta-d$ scattering length
is $\sim$ 4 fm, i.e. almost 6 times larger than the $\eta-$nucleon
scattering length \cite{Calen}. The present experimental 
results are consistent with the existence of a lightly bound state of 
$\eta-$mesons and $\alpha-$particles, although the
results remain somewhat inconclusive \cite{Wilkin,Tat}.

Theoretically the question of $\eta-^4$He bound states is
related to the question of the interaction between 
heavy quarkonia
and nuclei \cite{Brodsky,Manoh}. This issue has been 
investigated within the framework of the Skyrme model, 
which allows a coherent treatment
of the $\eta,\eta_c$ and $\eta_b$ mesons with the same
Lagrangian density in ref. \cite{Gobbi}.
The conclusion reached was that even
though the $\eta$ meson does not form bound states with
nucleons, the much heavier $\eta_c$ and $\eta_b$ mesons
are very likely to form such. Thus e.g. the $c\bar c$ analog
of the $\eta N$ $1/2^-$ resonance $N(1535)$ predicted
at $\simeq$ 2820 MeV would be stable against strong decay.

The Skyrme model has the conceptual advantage over more
phenomenological approaches to the coupling of $\eta$ mesons
and nuclei in that the same Lagrangian density with one
set of parameter values applies to several nuclear systems. The
model itself represents an effective representation of
the large $N_C$ limit of QCD \cite{Witten}.

To describe the coupling between the $\eta-$meson and
baryonic matter we employ Pari's version of the Skyrme
model \cite{Pari}:
\begin{equation}
{\cal L}=-{f_\pi^2 \over 4} tr\{L_\mu L^\mu\}
+{x\over 32e^2}tr\{[L_\mu,L_\nu]^2\}
+ {1-x\over 16e^2}\left[ (tr\{L_\mu L_\nu\})^2
-(tr\{L_\mu L_\mu\})^2 \right] + {\cal L}_{SB}
\end{equation}
Here $L_\mu\equiv U^\dagger \partial_\mu U$, where
$U$ is the $SU(3)$ extension of the skyrmion field,
$f_\pi$ is the pion decay constant and $e$ is the
inverse strength parameter of the quartic stabilizing
term. The term ${\cal L}_{SB}$ accounts for the chiral symmetry
breaking terms responsible for the different pseudoscalar
meson masses and decay constants. Its explicit form can be
found in e.g. ref.\cite{RS91}. The parameter $x$ determines 
the strength of the
coupling $1-x$ of the $\eta$ to baryonic matter. This
may be determined by the 
empirical value of the
real part of the $\eta-$nucleon
scattering length $a_{\eta N}=0.717\pm0.030$ \cite{Svarc}
(this value is close to that in ref.\cite{Green}).

The effective Lagrangian for the $\eta-$skyrmion system
can be obtained using the usual ansatz:
\begin{equation}
U=U_\pi e^{i\lambda_8\eta/f_\eta},
\end{equation}
where $U_\pi$ represents the SU(2) skyrmion field
and $\eta$ the field of the $\eta-$meson. For $U_\pi$ we
shall employ the rational map ansatz of ref.\cite{Houghton},
which provides a satisfactory description of the minimum
energy configurations of baryons with $B>1$. Thus
\begin{equation}
U_\pi=e^{i\vec\tau\cdot\hat\pi_n F(r)},
\end{equation}
where $F(r)$ is the chiral angle, which depends on the distance
to the center of the skyrmion, and $\hat\pi$ is defined
as the vector field
\begin{equation}
\hat\pi_n={1\over 1+|R_n(z)|^2}
\left( 
 2 Re \left[ R_n(z) \right], 2 Im \left[ R_n(z) \right],
1-|R_n(z)|^2
\right) .
\end{equation}
Here $R_n(z)$ is the rational map for winding (baryon) number
$B=n$, and the variable $z$ is defined terms of the usual 
spherical coordinates  $\theta, \varphi$ as $z\equiv 
tan(\theta/2)exp(i\varphi)$. For $n=1$, $R(z)=z$ and
(3) reduces
to Skyrme's hedgehog ansatz. 
The explicit forms of the rational maps for various values
of $n$ are given in ref.\cite{Houghton}. In the case of the
$\alpha-$particle, for which $n=4$, the
rational map takes the form
\begin{equation}
R_4(z)={z^4+2\sqrt{3}iz^2+1\over z^4-2\sqrt{3}iz^2+1}.
\end{equation}

Insertion of the ansatz (3),(4) into the Lagrangian density
of the model yields, upon expansion of (2) to second order
in the $\eta$ field, the following wave equation for the
$\eta-$meson in the skyrmion field:
\begin{equation}
\left[ 
{1\over r^2} {\partial\over \partial r}
\left( 
r^2\beta \ {\partial\over \partial r} 
\right)
+\alpha \ \omega_\eta^2 -\gamma^2 
\right] \eta(r)=0.
\end{equation}
where the $\eta$ meson has been assumed to be in an
$S$-state and the auxiliary functions $\alpha,\beta,\gamma$ have
been defined as
\begin{eqnarray}
\alpha &=& 1+{1-x\over e^2f_\eta^2}
\left[F^{'2}+2n{sin^2 F\over r^2}\right], \\
\beta  &=& 1+2n{1-x\over e^2f_\eta^2}{sin^2F\over r^2}, \\
\gamma^2 &=& {4 f_K^2 \over 3 f_\eta^2} m_K^2 - 
{m_\pi^2 f_\pi^2 \over 3 f_\eta^2} (2-cosF) .
\end{eqnarray}
Here, $\omega_\eta$ is the energy of the $\eta-$meson. The
differential equation (5) admits numerical solution for
any $n$, once the chiral angle $F(r)$ has been obtained by 
minimizing the corresponding multiskyrmion mass.

In our numerical calculations the following 3 sets of parameter values will
be used:
(1) $f_\pi=64.5$ MeV, $e=5.45$, $m_\pi=0$,
(2) $f_\pi=54$ MeV, $e=4.84$, $m_\pi=138$, 
(3) $f_\pi=93$ MeV, $e=4.25$, $m_\pi=138$. The
parameter sets (1) and (2) reproduce the empirical values
of the nucleon and $\Delta(1232)$ masses \cite{Zahed}, 
whereas set (3), which uses the empirical value for the pion decay
constant $f_\pi$, reproduces the empirical $N\Delta(1232)$
mass splitting, but overpredicts their absolute masses. The kaon mass
is taken at its empirical value $m_K = 495$ MeV and the
$\eta$ decay constant $f_\eta$ is determined so that
$f_\eta/f_\pi=1.28$, which ratio is consistent with the
empirical ratio 1.22 between the kaon and pion decay
constants \cite{Gobbi2}.

	The parameter $x$, which determines the strength of
the $\eta-$skyrmion coupling $1-x$ is determined so that
the continuum solution of (5) for $n=1$ yields the empirical 
value for the real part of the $\eta-N$ scattering length. 
With the parameter sets (1), (2) and (3)
above this value is obtained with $x=$0.92, 0.96 and
0.89 respectively. Given the value for $x$ the wave equation
(5) may be solved for increasing baryon number $n=A$ to
test for the existence of $\eta-$nucleus bound states. 

	In Table 1 we show the predicted binding energies
for $A=3-5$. 
While no bound state is formed by $\eta-$mesons and
deuterons (in conformity with experiment),
all parameter sets lead to the
existence of weakly bound states of the $\eta-$mesons and
$\alpha-$particles, with binding energies that vary from
30 KeV to 1.3 MeV.
With the parameter set (3) a bound state is also obtained
for $^3$He, but as the nucleon and $\Delta(1232)$ masses
are overestimated by several hundred MeV with that
parameter set, unless quantum corrections are taken into
account, this prediction appears as less reliable.
Consequently the binding energy estimate of 30 KeV 
for the $\eta-\alpha-$particle state also
appears more realistic than the value 1.3 MeV obtained with
parameter set 3.
In any case, the conclusion that there will be a bound state in the 
case of the $\alpha-$particle seems to be robust in view
of the fact that it appears for a wide range of parameter
values.

	One may speculate whether the $\eta-$meson coupling to
baryonic matter is strong enough to produce a bound Borromean
system of $\eta-$mesons and the otherwise unbound $^5$He.
A direct application of the present method of calculation
would lead to a binding energy of the $\eta-$meson to the
5-nucleon system of about 1 MeV (see Table 1). 

	The main conclusion is then that the $\eta-$meson should
form a bound state with the $\alpha-$particle, with a binding
energy of about 30 KeV. This prediction conforms with recent
experimental information on the $\eta-$nuclear interaction
\cite{Calen,Wilkin,Svarc}. Precision experiments at cooler
ring facilities are expected to settle the issue in the
near future.

\vspace{1.5cm}

This work was initiated while the authors were participating
at the joint JLAB-ECT$^*$ workshop on ``N$^*$ and nonpertubative QCD" at 
the ECT$^*$ in Trento. We thank the organizers
of the workshop and the staff of the ECT$^*$ for their  
hospitality during that period. The work of NNS is partially supported by 
grants of Fundaci\'on Antorchas and ANPCyT, Argentina and that
of DOR by the Academy of Finland under contract 34081.

\vspace{2cm}

\begin{center}
Table 1. The predicted binding energies $B_A$ for nuclei
with $A$=3,4 and 5
as obtained for the 3 sets of parameters described in the
text.
\vspace{1cm}

\begin{tabular}{|c|c|c|c|c|}\hline

  Set   &   $B_3$       &    $B_4$  &   $B_5$      \\ 
      &   MeV    &    MeV    &    MeV          \\ \hline
1  &         --     & 0.03  &  1.1       \\ \hline
2  &           --     & 0.03   &  0.5     \\ \hline
3  &          0.02    &  1.3   &  3.2     \\ \hline
\end{tabular}
\end{center}

\end {document}